\documentclass[prd,twocolumn,amssymb,12,showpacs]{revtex4-1}
\pdfoutput=1
\usepackage{graphicx} 
\usepackage{units}
\usepackage{xcolor}
\usepackage{amsmath}
\usepackage{amssymb}
\usepackage{stmaryrd}
\usepackage{esint}
\usepackage{filecontents}
\usepackage{siunitx}
\usepackage{comment}
\usepackage[bottom]{footmisc}
\usepackage{diagbox}
\usepackage[unicode=true,pdfusetitle,
 bookmarks=true,bookmarksnumbered=false,bookmarksopen=false,
 breaklinks=true,pdfborder={0 0 0},backref=true,colorlinks=true,linkcolor=blue!50!black,urlcolor=red!50!black,citecolor=blue!50!black]
 {hyperref}
\usepackage{breakurl}

\begin{document}
\title{Gravitational waves from inspiralling compact binaries in conformal gravity}
\author{Patric H\"olscher} \email{patric.hoelscher@physik.uni-bielefeld.de} \affiliation{Fakult\"at f\"ur Physik, Bielefeld University, Postfach 100131, 33501 Bielefeld, Germany} \author{Dominik J. Schwarz} \email{dschwarz@physik.uni-bielefeld.de} \affiliation{Fakult\"at f\"ur Physik, Bielefeld University, Postfach 100131, 33501 Bielefeld, Germany}

\begin{abstract}
We investigate the production of gravitational waves during the inspiral of compact 
binaries close to their merger in the context of a conformal gravity model. The model 
incorporates five massive polarization degrees of freedom, besides the two massless 
gravitational wave polarizations of general relativity. 
For small graviton mass, we find that the amplitude of the gravitational waves is strongly 
suppressed as compared to general relativity and decreases 
as coalescence is approached, which contradicts the observational fact. We conclude that
this model with small graviton mass, including a regime that can explain galaxy rotation 
curves without dark matter, cannot describe the observed gravitational wave events.
For a large graviton mass, the modifications to the waveform, compared to the one from 
general relativity, are negligible on the relevant distance scales and hence 
a conformal gravity model with a large graviton mass is in agreement with LIGO/VIRGO 
observations and leads to chirp mass and distance estimates that agree with 
those from general relativity.
\end{abstract}
\maketitle

\section{Introduction\label{sec:Introduction}}

With the detection of gravitational waves (GW) by the aLIGO and VIRGO observatories the era of GW astronomy started. Several binary black hole mergers were detected \citep{abbott2016observation,2016_Abbott_GW151226_Observationofgravitationalwavesfroma22-solar-massbinaryblackholecoalescence_PRL,2017_Scientific_GW170104_Observationofa50-Solar-MassBinaryBlackHoleCoalescenceatRedshift0.2_PRL,2017_Abbott_GW170608Observationofa19SolarMassBinaryBlackHoleCoalescence_TAJL,2017_Abbott_GW170814_Athree-detectorobservationofgravitationalwavesfromabinaryblackholecoalescence_PRL,2018AbbottGWTC1Gravitational2}. 
The analysis of these events shows that the predictions of general relativity (GR) are in very 
accurate agreement 
with these observations, which at the same time open a fantastic window of opportunity 
to test the physics of gravity beyond GR. 
Recently also the first binary neutron star merger (GW170817) has been detected 
with electromagnetic follow-up measurements in nearly the whole electromagnetic spectrum 
coming from GRB 170817A
\citep{2017_Abbott_GW170817_PRL,2017_Abbott_Multi-messengerobservationsofabinaryneutronstarmerger_AJL,2017_Coulter_SwopeSupernovaSurvey2017aSSS17atheOpticalCounterparttoaGravitationalWaveSource_S}. 
From the measured difference in the arrival time of the gravitational and electromagnetic signal strong constraints on the speed of the GWs are given. This rules out several theories of modified gravity
\citep{2017_Abbott_GravitationalWavesandGammaRaysfromaBinary,
2017_Lombriser_ChallengestoSelfAccelerationinModifiedGravity,
2017_Baker_StrongConstraintsonCosmologicalGravityfromGW170817andGRB170817A_Prl,2017_Creminelli_DarkEnergyAfterGW170817andGRB170817A_PRL,2017_Ezquiaga_DarkEnergyAfterGW170817DeadEndsandtheRoadAhead_PRL,
2017_Sakstein_ImplicationsoftheNeutronStarMergerGW170817forCosmologicalScalarTensorTheories_Prl,
2018_Nersisyan_GravitationalWaveSpeedImplicationsforModelswithoutaMassScale_,
2018_Akrami_NeutronStarMergerGW170817StronglyConstrainsDoublyCoupledBigravity_apa}, especially when the graviton has a small mass. 

In this work we analyze the wave form of GWs in conformal gravity (CG) to investigate if the GW detections of aLIGO and VIRGO can be explained within this theory. 
It is special because it is based on a Weyl invariant action, which is unique 
up to the choice of the matter content, the coupling constants and their signs. 
Explicit mass scales are forbidden by this symmetry, and become manifest only after fixing Weyl 
symmetry. The Einstein-Hilbert term and the cosmological constant appear then together 
with all other masses and the CG models exhibit two gravitational modes, 
a massless and a massive one. Thus these models cannot be constrained based on 
the signal travel time alone, as the massless mode 
behaves as in GR and a more detailed analysis of the wave forms is required.

A special case is Mannheim's CG model for which an exact solution for the 
gravitational potential for a static, spherically symmetric system was found 
\citep{1989MannheimKazanasExactvacuumsolutionTAJ}. In addition to the $1/r$-term in the 
Schwarzschild solution it contains a term linear in $r$, which can be used to fit a large number of 
galaxy rotation curves without the need for dark matter
\citep{1997MannheimAregalacticrotationTAJ,2013MannheimOBrienGalacticrotationcurves,mannheim2012fitting}. However, 
cosmic structure formation has not been analyzed yet.

The model of CG studied in this work is a fourth-order derivative theory. 
This makes them 
power-counting renormalizable \citep{1977StelleRenormalizationhigherderivativePRD,2016FariaQuantummassiveconformalTEPJC}, which means that they are 
candidates for a theory of quantum gravity.  But CG models 
with higher derivatives suffer from 
the Weyl ghost, although it is claimed that the ghost issue can be solved
\citep{2000MannheimDavidsonFourthordertheoriesaph,2007BenderMakingsensenonRoPiP,2008BenderMannheimExactlysolvablePPRD,2008BenderMannheimGivingghostJoPAMaT,2008BenderMannheimNoghosttheoremPRL,2013MannheimPTsymmetryasPTotRSoLAMPaES,2018_Mannheim_AntilinearityRatherThanapa,2016MannheimExtensionCPTtheoremPLB}.

In \citep{2018_Caprini_AstrophysicalGravitationalWavesinConformalGravity_apa} the authors have analyzed the gravitational radiation of stellar binary systems in a CG model (CGM) 
that inlcudes Mannheim's CG as a special case. A neutron star-white dwarf system in slow stationary inspiral on circular orbits in the Newtonian limit was studied and two parameter regimes, corresponding to a small and a large graviton mass, were identified. It was shown that for the case of a small graviton mass the radiated power cannot explain the decrease of the orbital period. However, it turned out there is a parameter regime with a large graviton mass which is in agreement with the observations and which contains a GR limit. 

For a small graviton mass the energy lost by the binary system could be stored in the near field 
by some mechanism, which means that this parameter regime is not necessarily ruled out. 
In the present work we use the measurements of GWs in the late phase of the evolution of a 
binary system. aLIGO and VIRGO measured GW signals consistent with the predictions of GR.
Hence, we investigate if it is possible to also fit the observed GW signal with wave forms predicted 
in the CGM. As we show below, this turns out to be impossible and we can thus rule out 
small graviton masses and therefore also Mannheim's model of CG.

In Sec. \ref{Conformal Gravity} we give an introduction to the CGM. In Sec. \ref{Gravitational Waves} we recap the results from \cite{2018_Caprini_AstrophysicalGravitationalWavesinConformalGravity_apa}, where the linearized equations of motion and their GW solutions for a binary system on circular orbits and in the Newtonian limit were worked out. We discuss the modifications to Kepler's third law in Sec. \ref{Kepler} and calculate the wave form in the late inspiral phase and compare it to GR. Lastly, we discuss our findings and conclude.

In the first two sections and in the beginning of Sec.~III we use natural units, 
hence $c = \hbar = 1$. Greek indices take values 
$0$ to $3$, whereas latin indices run from $1$ to $3$. Other conventions are specified 
in Appendix \ref{appendix A}.

\section{Conformal Gravity Models \label{Conformal Gravity}}

The action of models of CG is given by
\begin{align}
I & = - \alpha_{g}\int d^{4}x\sqrt{-g}C_{\lambda\mu\nu\kappa}C^{\lambda\mu\nu\kappa}+I_{M}
	\label{eq: Weyl action}
\end{align}
where $\alpha_{g}$ is a dimensionless coupling constant,
$x$ denotes the spacetime coordinates, $g$ 
is the determinant of the metric tensor $g_{\mu\nu}$,
$C_{\lambda\mu\nu\kappa}$ is the Weyl tensor and $I_M$ denotes the matter action. 
Using the Gauss-Bonnet term, 
which is a total derivative in four space-time dimensions 
and hence does not contribute to the field equations for the metric, 
we can rewrite \eqref{eq: Weyl action} as
\begin{align}
I = - 2 \alpha_{g}\int d^{4}x\sqrt{-g}
\left(R_{\mu\kappa}R^{\mu\kappa} - \frac{1}{3}R^{2}\right) 
+ I_{M},
 	\label{eq:Weyl_action_gauss_bonnet}
\end{align}
where $R_{\mu\kappa}$ and $R$ are the Ricci tensor and scalar.
The gravitational action is invariant under Weyl transformations of the metric,
\begin{equation}
g_{\mu\nu}\left(x\right)\rightarrow\Omega^{2}\left(x\right)g_{\mu\nu}\left(x\right),
	\label{eq: Weyl transformation}
\end{equation}
where the conformal factor $\Omega$ is real, positive and smooth.  

Variation of the action \eqref{eq:Weyl_action_gauss_bonnet} with respect to $g_{\mu\nu}$ leads to the equations of motion for the gravitational field \citep{1921BachZurWeylschenrelativitatstheorieMZ}
\begin{equation}
4 \alpha_{g} W^{\mu\nu} = 4\alpha_{g} \left[2C_{\;\;\;;\lambda;\kappa}^{\mu\lambda\nu\kappa}-C^{\mu\lambda\nu\kappa}R_{\lambda\kappa}\right] = T_{M}^{\mu\nu},
	\label{Bach equation}
\end{equation}
where $W_{\mu\nu}$ is the Bach tensor and 
\begin{equation}
T_{M}^{\mu\nu} \equiv \frac{2}{(-g)^{1/2}} \frac{\delta I_M}{\delta g_{\mu\nu}}
\end{equation}
is the matter energy-momentum tensor.

The matter energy-momentum tensor should also be Weyl invariant. Hence, the most general local matter action for a  generic scalar and a spinor field coupled conformally to gravity is
\citep{1990MannheimConformalcosmologynoGRaG}
\begin{widetext}
\begin{equation}
I_{M} = -\int d^{4}x\sqrt{-g}\left[\epsilon\left(-\frac{S^{,\mu}S_{,\mu}}{2} + \frac{S^{2}R}{12}\right)+\lambda S^{4} + i\bar{\psi}\gamma^{\mu}\left(x\right)\left[\partial_{\mu} + \Gamma_{\mu}\left(x\right)\right]\psi - \xi S\bar{\psi}\psi\right]. 
	\label{eq: matter action}
\end{equation}
\end{widetext}
$S(x)$ represents a self-interacting real scalar field and $\psi(x)$ is a generic 
spin-$\nicefrac{1}{2}$ fermion field.
$\xi$ and $\lambda$ are dimensionless coupling constants, $\gamma^{\mu}(x)$ are the 
vierbein-dependent Dirac-gamma matrices, $\bar{\psi} = \psi^{\dagger} \gamma^{0}$ and 
$\Gamma_{\mu}(x)$ is the fermion spin connection \cite{simplifiednotation}. To be invariant under local 
Weyl transformations the several fields have to transform as 
$S(x) \rightarrow \Omega^{-1}(x) S(x)$,
$\psi(x) \rightarrow \Omega^{-3/2}(x) \psi(x)$
and $g_{\mu\nu}(x) \rightarrow \Omega^{2}(x) g_{\mu\nu}(x)$. 
The exponent of the conformal factor is called conformal weight.

In the action \eqref{eq: matter action} only the combination of the two terms in 
round brackets is Weyl invariant. Hence, we introduce the parameter $\epsilon$, 
which can take the values $-1$ or $+1$. In the first case, the theory corresponds to the 
model advocated by Mannheim and Kazanas 
\cite{1989MannheimKazanasExactvacuumsolutionTAJ}. The case $\epsilon = + 1$ was
called massive conformal gravity in our previous work
\cite{2018_Caprini_AstrophysicalGravitationalWavesinConformalGravity_apa}, 
but it was pointed out in \cite{faria2019comment} that this naming is in conflict with previous 
terminology. In order to avoid any further confusion, we refer to both cases as the same 
conformal gravity model (referred to as CGM below) which has two distinct regimes specified 
by $\epsilon = \pm 1$.

Since the action $I$ 
is Weyl invariant, it is always possible to choose a frame in which the scalar field is constant 
\begin{equation}
S(x)\rightarrow S'(x) = \Omega^{-1}(x) S(x) = S_{0} = \text{const.},
\end{equation}
with $\Omega(x) = S(x)/S_{0}$.
This is called the Higgs  or unitary gauge
\citep{horne2016conformal,faria2014massive}. In order to make connection to
GR, one chooses
\begin{subequations} 
\begin{align}
8 \pi G & \equiv \frac{6}{S_{0}^{2}}, 
	\label{eq:relation to GR-1}\\
\Lambda & \equiv 6 \lambda S_{0}^{2},
	\label{cosm const} 
\end{align}
\end{subequations}
where $G$ denotes Newton's constant and $\Lambda$ is interpreted as the cosmological constant. 
This is a reasonable choice because for $\epsilon = +1$ and in the limit $\alpha_g \rightarrow 0$, 
the action $I$ becomes the Einstein-Hilbert action with minimally coupled matter.

Variation of the action \eqref{eq: matter action} with respect to $g_{\mu\nu}$ leads to the matter energy-momentum tensor 
\begin{equation}
T_{\mu\nu}^{M} = T_{\mu\nu} + \frac{\epsilon}{8\pi G}\left(R_{\mu\nu} - \frac{1}{2}g_{\mu\nu}R\right) - g_{\mu\nu}\frac{\Lambda}{8\pi G},
\end{equation}
where 
\begin{equation}
T_{\mu\nu} = \frac{1}{2} \left[ 
i\bar{\psi}\gamma_{\mu}(x)\left[\partial_{\nu}+\Gamma_{\nu}(x)\right]\psi + 
(\mu \leftrightarrow \nu) \right]
	\label{emt fermion}
\end{equation}
is the fermionic energy-momentum tensor.

The equation for the gravitational field now becomes \citep{1985SchmidtstaticsphericalsymmetricAN,1985_Schmidt_SolutionsoftheLinearizedBachEinsteinEquationintheStaticSphericallySymmetricCase_AN}, 
\begin{equation}
4\alpha_g W_{\mu\nu} = T_{\mu\nu} + 
\frac{1}{8\pi G} \left[ \epsilon G_{\mu\nu} - g_{\mu\nu} \Lambda \right],
	\label{eq: energy-momentum tensor in higgs gauge-1}
\end{equation}
where $G_{\mu\nu}$ denotes the Einstein tensor.
From the trace of \eqref{eq: energy-momentum tensor in higgs gauge-1} we get
\begin{equation}
\epsilon R + 4 \Lambda = 8 \pi G T,
\end{equation}
where $T$ is the trace of the fermionic matter energy-momentum tensor.
Note that the fermion energy-momentum tensor 
is covariantly conserved,
\begin{equation}
T_{\:\:;\nu}^{\mu\nu} = 0,
\end{equation}
due to the Bianchi identities for the Bach and Einstein tensors.

We observe that it is convenient to introduce a new parameter $m_g$ with the dimensions 
of a mass,
\begin{equation}
m_g^2 \equiv  \frac{1}{32 \pi G \alpha_g}.
\end{equation}
We can then write 
\begin{equation}
 - \epsilon G_{\mu\nu} + g_{\mu\nu} \Lambda + \frac{1}{m_g^2} W_{\mu\nu} = 8\pi G T_{\mu\nu},
 	\label{eq: Modified Bach equation}
\end{equation}
and observe that in the limit $m_g \to \infty$, the Einstein equations are recovered for  
$\epsilon = +1$. 
Mannheim's case ($\epsilon = -1$) does not contain general relativity as a limit.

\section{Gravitational Waves \label{Gravitational Waves}}

In \citep{2018_Caprini_AstrophysicalGravitationalWavesinConformalGravity_apa} we linearized 
the equations of motion for the metric and derived the gravitational wave solutions for 
conformal gravity with both signs of $\epsilon$ and for different parameter regions.
In a flat Minkowski background we found a massless and a massive mode of the metric perturbation given by
\begin{equation}
h_{\mu\nu} = \epsilon\left( H_{\mu\nu} + \Psi_{\mu\nu} \right),
	\label{metric split}
\end{equation}
where 
\begin{align}
\Psi_{\mu\nu} &= \frac{1}{m_g^2} \left(\Box h_{\mu\nu} - \frac{1}{3}\eta_{\mu\nu}R\right)
	\label{psi}
\end{align}
represents the massive mode and $H_{\mu\nu}$ is the massless mode.
Using \eqref{metric split} and \eqref{psi} in the linearized version of \eqref{eq: Modified Bach equation} leads to
\begin{subequations}
\begin{align}
\Box\bar{H}_{\mu\nu} &= -16\pi G T_{\mu\nu},\;\; \partial^\mu \bar{H}_{\mu\nu} = 0,
	\label{massless_eom}\\
\left(\Box-\epsilon m_{g}^{2}\right)\hat{\Psi}_{\mu\nu} 
&= 
16\pi G T_{\mu\nu},\;\;\partial_{\rho}\partial_{\sigma}\hat{\Psi}^{\rho\sigma} = 0 
	\label{massive_eom_hat},
\end{align}
\end{subequations}
where $\bar{H}_{\mu\nu} \equiv H_{\mu\nu} - 1/2 \eta_{\mu\nu}H$ and $\hat{\Psi}_{\mu\nu} \equiv \Psi_{\mu\nu}-\eta_{\mu\nu}\Psi$. 
This result holds in the Teyssandier gauge 
\begin{equation}
Z_{\mu} = - m_{g}^{-2}
\left[\left(\Box-\epsilon m_{g}^{2}\right)\partial_{\rho}\bar{h}_{\mu}^{\rho} + (1/3) \partial_{\mu}R\right] = 0,
\end{equation}
see \citep{1989_Teyssandier, 2017_Accioly_Interestingfeaturesofageneralclassofhigherderivativetheoriesofquantumgravity}.
It can be shown that there are two massless polarization degrees of freedom encoded 
in $H_{\mu\nu}$ and five massive ones encoded in $\Psi_{\mu\nu}$. The solutions for 
the massless modes are identical to the case of general relativity. For the massive modes we found 
that binary systems can excite only two of the five modes and, as in general relativity, 
the gravitational waves are sourced by time varying quadrupole moments of the mass distribution.


The solutions for the metric perturbation for a binary system of two compact objects of mass $m_1$ and $m_2$ in circular motion in the Newtonian limit and in the center of mass frame were derived in \citep{2018_Caprini_AstrophysicalGravitationalWavesinConformalGravity_apa}. The dynamics of the system can be reduced to a one-body problem with reduced mass $\mu = m_1 m_2/(m_1 + m_2)$. The relative coordinates for a circular path in the xy-plane are given by
\begin{subequations}
\begin{align}
x(t) &= -R \sin\left(\omega_\mathrm{s} t \right),\\
y(t) &= R \cos\left(\omega_\mathrm{s} t \right),
\end{align}
\end{subequations}
where $R$ is the distance between the two compact objects and $\omega_s$ is the orbital 
frequency. The corresponding circular frequency of the emitted gravitational waves is 
$\omega_\mathrm{gw} = 2 \omega_\mathrm{s}$.
To find an approximate analytic solutions to Eq.~\eqref{massive_eom_hat} we distinguish two cases.
The relation between the graviton mass $m_g$ and the frequency $\omega_s$ determines 
the physical behavior of the massive mode.

For a small graviton mass, $m_g c^2 < \hbar \omega_\mathrm{gw}$, in \citep{2018_Caprini_AstrophysicalGravitationalWavesinConformalGravity_apa} we found
\begin{subequations}
\begin{align}
&h_{11}(t,r)  = -h_{22}(t,r) \nonumber \\
&= \frac{\epsilon}{r}\frac{G\mu R^{2}\omega_\mathrm{gw}^{2}}{c^4} \left[ \cos\left(\omega_\mathrm{gw}t_0 \right) - \cos\left(\omega_\mathrm{gw}t_\mathrm{m}\right) \right],
	\label{solution case1/3 comp 11}\\
&h_{12}(t,r) = h_{21}(t,r) \nonumber \\
&= \frac{\epsilon}{r}\frac{G\mu R^{2}\omega_\mathrm{gw}^{2}}{c^4} \left[\sin\left(\omega_\mathrm{gw}t_0 \right) - \sin\left(\omega_\mathrm{gw}t_\mathrm{m}\right)\right],
	\label{solution case1/3 comp 12}
\end{align}
\end{subequations}
where now $t$ denotes the time of an observer at distance $r$ to the source, 
$t_0 = t - r/c$ is the retarded time for the massless mode and 
$t_{m} = t - v_{g,\epsilon}r/c^2$ is the retarded time for the massive mode.
$v_{g,\epsilon} =  c [1-\epsilon m_g^2 c^4/\hbar^2\omega_\mathrm{gw}^2]^{1/2}$ is the group velocity of the massive gravitational wave.

For $\epsilon = + 1$ and $m_g c^2 > \hbar \omega_\mathrm{gw}$, in \citep{2018_Caprini_AstrophysicalGravitationalWavesinConformalGravity_apa} we found
\begin{subequations}
\begin{align}
&h_{11}\left(t,r\right) = -h_{22}(t,r) \nonumber \\
&= \frac{1}{r} \frac{G \mu R^{2}\omega_\mathrm{gw}^{2}}{c^4}\left[\cos(\omega_\mathrm{gw}t_0)-e^{-k_{\omega}r}\cos(\omega_\mathrm{gw}t)\right],			
	\label{eq: solution of case 2}\\
&h_{12}(t,r) = h_{21}(t,r)  \nonumber \\ 
& = \frac{1}{r} \frac{G \mu R^{2}\omega_\mathrm{gw}^{2}}{c^4}\left[\sin(\omega_\mathrm{gw}t_0)-e^{-k_{\omega}r}\sin(\omega_\mathrm{gw}t)\right],
	\label{eq: solution of case 2_2}
\end{align}
\end{subequations}
where $k_\omega = \sqrt{(m_g c/\hbar)^2 - (\omega_{\mathrm{gw}}/c)^2}$. This is the GR solution modified by an exponentially damped term.
We do not consider the case of a large graviton mass for $\epsilon = -1$, because in the Newtonian limit this model leads to repulsive gravity.

\section{Kepler's Third Law and the Wave Form of inspiralling binaries \label{Kepler}}

To investigate how gravitational radiation influences the orbits of the binary system, 
we have to find Kepler's third law in the CGM. For the analysis we consider an idealized 
system of two compact objects of 
mass $m_1$ and $m_2$ in the Newtonian limit and in the center of mass frame with 
reduced mass $\mu$ and total mass $M = m_1 + m_2$. Assuming that the 
Newtonian approximation is applicable in the stationary and quasicircular phase 
of the merger, we can use the Newtonian potential energy. To find Kepler's 
third law it is necessary to discuss on which scales the modifications to the $1/R$-potential 
become important. 

For the following estimates we consider the typical gravitational waves 
as observed with the LIGO/VIRGO detectors. These gravitational waves are in the audio band, 
thus have a typical frequency of $1$ kHz and a typical wavelength of $300$ km. 

Let us first consider $\epsilon = + 1$, for which the Newtonian potential is modified by 
a Yukawa potential. From the analysis of the GWs we found that we have to distinguish two 
regimes. The case of a large graviton mass $m_g \gg  \hbar 
\omega_\mathrm{gw}/c^2$ applies for the reduced Compton wavelength $\lambdabar_g \equiv \hbar /m_g c \ll c/\omega_\mathrm{gw} \approx 50$ km.
The Newtonian potential of conformal gravity is then given as
\begin{align}
E_{\mathrm{pot},\epsilon = + 1} & = -\frac{G \mu M}{R}\left( 1 - \frac{4}{3} e^{-R/\lambdabar_g}\right).
	\label{grav potential MCG}
\end{align}
Tests of the inverse square law from terrestrial to sub-mm distances lead to the 
constraint  $\lambdabar_g < \SI{d-5}{m}$  \cite{2009AdelbergerGundlachHeckelHoedlSchlammingerTorsionbalanceexperimentsPiPaNP}
(corresponding to $m_g > \SI{d-38}{kg}$ and frequencies $f_g > \SI{4.8d12}{Hz}$).

For small graviton mass $m_g \ll \hbar \omega_{\mathrm{gw}}/c^2$,  
the reduced Compton wavelengths are much longer than $50$ km and the modified 
Newtonian potential must read
\begin{align}
E_{\mathrm{pot},\epsilon = + 1} & = -\frac{4 G \mu M}{R}\left(e^{-R/\lambdabar_g} - \frac{3}{4} \right).
	\label{grav potential MCG_smallmass}
\end{align}
Modifications on large length scales are constrained by terrestrial and Solar System tests of 
the inverse square law, which lead to $\lambdabar_g > \SI{d16}{m} \approx 0.3$ pc 
\cite{2009AdelbergerGundlachHeckelHoedlSchlammingerTorsionbalanceexperimentsPiPaNP}
(corresponding to $m_g < \SI{d-58}{kg}$ and $f_g  < \SI{4.8d-9}{Hz}$).

For $\epsilon = -1$, 
\begin{align}
E_{\mathrm{pot},\epsilon = -1} & = -\frac{G \mu M}{R} + \frac{\gamma_{\odot} \mu M }{2} c^2 R,
	\label{eq:grav_pot_CG}
\end{align}
where $\gamma_\odot$ is an integration constant with dimension of inverse mass times 
inverse distance. 

The term linear in $R$ in Eq.~\eqref{eq:grav_pot_CG} was used to fit galaxy 
rotation curves without dark matter leading to $\gamma_\odot = \SI{2.4d-69}{kg^{-1}m^{-1}}$ 
\cite{mannheim2012fitting}. This corresponds to typical length scale $\lambdabar = (4 G/\gamma_\odot c^2)^{1/2} \approx 36$ kpc and therefore contributes noticeably to the potential 
energy on galactic distance scales $(R > \SI{1}{kpc})$, but is negligible on Solar System scales.

Assuming that the semimajor axis of the binary systems is well below galactic distance 
scales (much smaller than the parsec scale) and well above the sub-mm scale, 
the modifications in \eqref{grav potential MCG}, \eqref{grav potential MCG_smallmass} and 
\eqref{eq:grav_pot_CG} to the $1/R$-term are negligible.	
Hence, Kepler's third law for circular orbits is approximately given by 
\begin{equation}
\omega_s^2 \approx \frac{GM}{R^3},
	\label{Kepler CG}
\end{equation}
and for the orbital energy we find
\begin{align}
E_{\mathrm{orbit}} &= E_{\mathrm{kin}} + E_{\mathrm{pot}}\nonumber\\
&\approx -\frac{G\mu M}{2R}
	\label{orbital energy}
\end{align}
as in the Newtonian approximation of GR.
Inserting \eqref{Kepler CG} into \eqref{orbital energy} we get
\begin{align}
E_{\mathrm{orbit}} \approx -\left( \frac{G^2 M_c^5 \omega^2_{\mathrm{gw}}}{32} \right)^{1/3},
\end{align}
where $\omega_{\mathrm{gw}} = 2\omega_s$ and $M_c = \mu^{3/5}M^{2/5}$ is the chirp mass.
The energy loss of the system is given by
\begin{align}
\dot{E}_{\mathrm{orbit}} \approx -\frac{2}{3} \left(\frac{G^2 M_c^5}{32\omega_{\mathrm{gw}}} \right)^{1/3} \dot{\omega}_{\mathrm{gw}}.	
	\label{energy loss}
\end{align}

\subsection{Large Graviton Mass \label{Massive Conformal Gravity}}

In \citep{1999BarabashShtanovNewtonianlimitconformalPRD,2007BarabashPyatkovskaWeakfieldlimitapa} 
it is pointed out that the CGM with a large graviton mass cannot fit galaxy rotation curves without dark matter, as can also be seen most easily from Eq.~(\ref{grav potential MCG}). Nevertheless,  
the large graviton mass case is very interesting because of its GR limit.

The massive part of the gravitational wave is damped and effectively only the massless graviton, which is the same as in GR, contributes. Nevertheless, there is a profound difference to GR, since this theory is power-counting renormalizable \citep{1977StelleRenormalizationhigherderivativePRD,2016FariaQuantummassiveconformalTEPJC} and hence could provide a viable theory of quantum gravity.

In this case the Newtonian potential is given by Eq.~\eqref{grav potential MCG}.
This means that the Yukawa term in \eqref{grav potential MCG} becomes important only on sub-mm distance scales. For binary systems in the inspiral phase the distance between the objects is always macroscopic (at least larger than two Schwarzschild radii) and hence we can 
neglect this term for the analysis of gravitational radiation from macroscopic binary systems. 

The radiated power is given by
\citep{2018_Caprini_AstrophysicalGravitationalWavesinConformalGravity_apa}
\begin{align}
P = P_{GR} = \frac{32 c^5}{5G} \left( \frac{G M_c \omega_{\mathrm{gw}}}{2 c^3} \right)^{10/3}.
\end{align}
In the following we go beyond the quasistationary case that we have considered before.

Equating $\dot{E}_{\mathrm{orbit}} = -P$ and solving for $\dot{\omega}_{\mathrm{gw}}$ yields
\begin{equation}
\dot{\omega}_{\mathrm{gw}} = \frac{12}{5} 2^{1/3} \left(\frac{G M_c}{c^3} \right)^{5/3} \omega_{\mathrm{gw}}^{11/3}.
	\label{diff eq frequency GR}
\end{equation}
Integrating \eqref{diff eq frequency GR} we get 
\begin{equation}
\omega_{\mathrm{gw}}(\tau) = \frac 14  \left( \frac{5}{\tau}\right)^{3/8} 
\left( \frac{G M_c}{c^3} \right)^{- 5/8},
\end{equation}
where $\tau = (t_{\mathrm{coal}} - r/c) - (t - r/c) = t_{\mathrm{coal}} - t$ is the time 
to coalescence as measured by a signal that propagates with the speed of light.

\begin{figure}
\includegraphics[width = \linewidth]{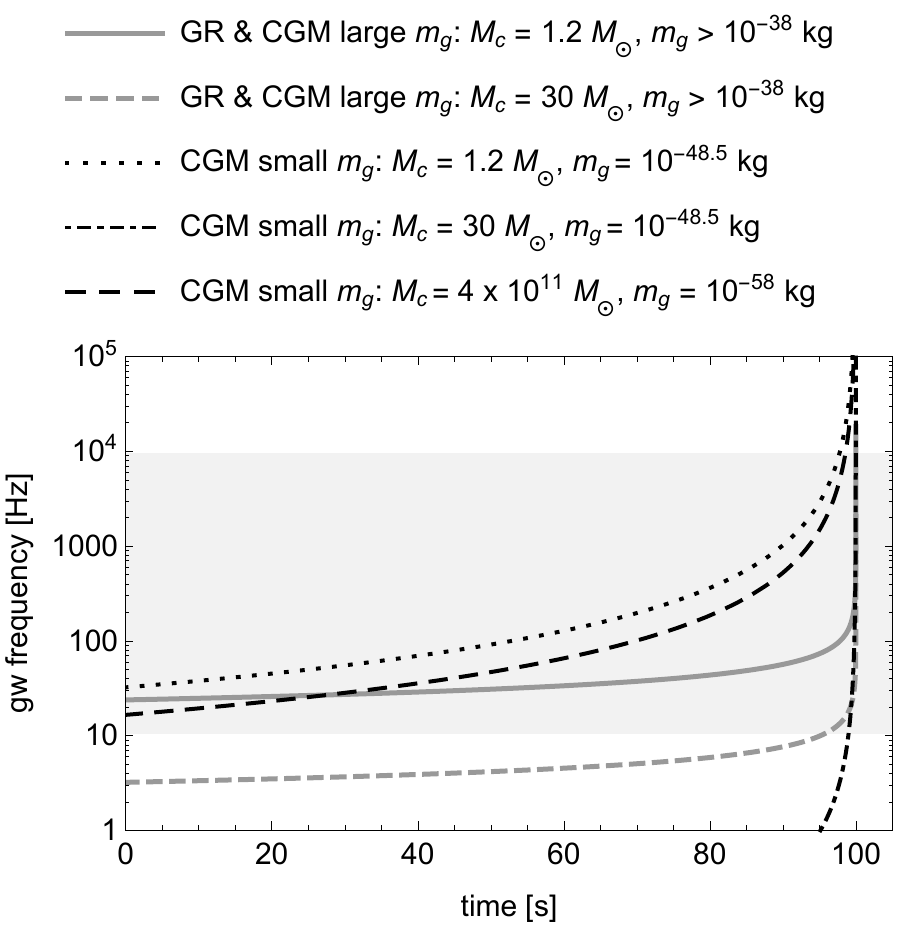}
\caption{Time evolution of the gravitational wave frequency $f_\mathrm{gw}$ of massless 
modes during the last 100 seconds before coalescence. 
The shaded region indicates the LIGO/VIRGO frequency band. The grey lines show the 
GR prediction, which is identical to the CGM prediction with a large graviton mass. 
The black lines show three examples for the CGM with a small graviton mass. 
Solar system tests of gravity imply that $m_g < 10^{-58}$ kg. \label{fig:frequencies}}
\end{figure}

Figure \ref{fig:frequencies} shows how the frequency of a gravitational wave, $f_{\mathrm{gw}} = \omega_{\mathrm{gw}}/2\pi$, evolves over 
the last 100 seconds before coalescence (in the approximation used in this work). Our 
approximation breaks down at frequencies above the LIGO/VIRGO 
frequency band (10 Hz to 10 kHz). To give two examples for the CGM with a 
large graviton mass 
we show the evolution of gravitational wave frequency for chirp masses of $1.2 M_\odot$, corresponding to a system of two neutron stars with masses of $1.4 M_\odot$ each, and $30 M_\odot$, a value typical for a pair of black holes.

We look at the wave form of the GWs produced by an object in quasicircular motion which is on an orbit in the $xy$-plane
\begin{subequations}
\begin{align}
x(\tau) &= R(\tau)\sin\left(\frac{\Phi(\tau)}{2}\right),\\
y(\tau) &= R(\tau)\cos\left(\frac{\Phi(\tau)}{2}\right).
\end{align}
\end{subequations}
Since the orbit is not stationary, we have to do the following replacements: $R \rightarrow R(\tau)$, which leads to terms proportional to $\dot{R}$ in the GW solutions when calculating the time derivatives of the quadrupole moment,
but as long as $\dot{\omega}_s \ll \omega_s^2$ we can neglect radial velocities. Using \eqref{diff eq frequency GR} this condition translates to $\omega_{\mathrm{gw}} \ll 0.51 (c^3/G M_c) $. This means we can neglect $\dot{R}$ as long as $f_{\mathrm{gw}} \ll \SI{16.6}{kHz} (M_\odot/M_c)$. 

Additionally, we have to replace $\omega_{\mathrm{gw}}$ with $\omega_{\mathrm{gw}}(\tau)$ in the amplitude of \eqref{eq: solution of case 2} -- \eqref{eq: solution of case 2_2} and 
$\omega_{\mathrm{gw}} t_0$ becomes $\Phi(\tau)$ in the argument of the trigonometric functions. 
The phase is defined as the circular frequency of the GW integrated over time
\begin{equation}
\Phi(\tau) = \int_{\tau_i}^\tau d\tau'\, \omega_{\mathrm{gw}}(\tau') + \Phi_i,
\end{equation}
where $\Phi_i$ is the phase at some initial $\tau_i$.
This leads to 
\begin{equation}
\Phi(\tau) =  \Phi_i - 2 \left( \frac{GM_c}{c^3} \right)^{\!\! -5/8} 
\left[ \left(\frac{\tau_i}{5}\right)^{\! 5/8} \!\! - \left(\frac{\tau}{5}\right)^{\!5/8} \right].
\end{equation}

Finally, we can find the wave form in terms of the time to coalescence 
$\tau$,
\begin{subequations}
\begin{align}
h_{11}(t,r)  = -h_{22}(t,r) &= \frac{c}{r}\left(\frac{GM_c}{c^3}\right)^{\!\! 5/4}\!\!\left(\frac{5}{\tau}\right)^{\!\! 1/4}\!\! \cos[\Phi(\tau)], 
	\label{solution large mass 11}\\	 
h_{12}(t,r) = h_{21}(t,r) &= \frac{c}{r}\left(\frac{GM_c}{c^3}\right)^{\!\! 5/4}\!\! 
\left(\frac{5}{\tau}\right)^{\!\! 1/4}\!\! \sin[\Phi(\tau)].
	\label{solution large mass 12}
\end{align}
\end{subequations}
This is the same result as in GR and the predictions are consistent with the observed GW events. Hence, it leads to the same predictions on the chirp mass and the distance to the source.

\subsection{Small Graviton Mass \label{conformal gravity}}

For the case of a small graviton mass $(m_{g} < \hbar \omega_{\mathrm{gw}}/c^2)$,
the power radiated into GWs is given by
\citep{2018_Caprini_AstrophysicalGravitationalWavesinConformalGravity_apa}
\begin{equation}
P = \frac 12 \left(\frac{m_{g} c^2}{\hbar \omega_{\mathrm{gw}}}\right)^{\! 2} \frac{32 c^5}{5G} \left( \frac{G M_c \omega_{\mathrm{gw}}}{2 c^3} \right)^{10/3}.
	\label{radiatedpower}
\end{equation}
We now go beyond the quasistationary limit by
equating $\dot{E}_{\mathrm{orbit}} = -P$ and solving for $\dot{\omega}_{\mathrm{gw}}$ yields
\begin{align}
\dot{\omega}_{\mathrm{gw}} &= \frac{6}{5} 2^{1/3} \left(\frac{m_g c^2}{\hbar}\right)^{\!\!2} \!\! \left(\frac{G M_c}{c^3} \right)^{\!\! 5/3} \omega_{\mathrm{gw}}^{5/3}.
\label{eqn:omegadotsmallmass}
\end{align}
We follow the same procedure as above and find with
$\tau = t_\mathrm{coal} - t$ being the time interval to coalescence as measured by a 
fiducial observer at the position of the source,
\begin{align}
\omega_{\mathrm{gw}} (\tau) 
&= \left(\frac{m_{g} c^2}{\hbar}\right)^{-3} \frac{1}{32} \left(\frac{5}{\tau}\right)^{3/2} \left(\frac{GM_c}{c^3} \right)^{-5/2}.
	\label{frequencyGW}
\end{align}
Note that $\omega_{\mathrm{gw}}$ diverges at $t = t_{\mathrm{coal}}$. This is no problem, 
since the two objects merge before the divergence takes place and further the Newtonian approximation breaks down, when the two objects come to close together.
Inserting numerical values in \eqref{frequencyGW} we find
\begin{equation}
f_{\mathrm{gw}}
= \SI{1.67d33}{Hz}\! \left( \frac{\SI{d-58}{kg}}{m_g} \right)^{\!\!3} \!\! \left( \frac{M_\odot}{M_c} \right)^{\!\!5/2}\!\!\!\left( \frac{\SI{1}{s}}{\tau} \right)^{\!\!3/2} \!\!\!.
\end{equation}

The evolution of the gravitational wave frequency towards coalescence is shown in 
Fig.~\ref{fig:frequencies}. The dotted, dash-dotted and long dashed lines show the case of 
the CGM with 
small graviton mass. For the dotted and dash-dotted line we have tuned the gravitational 
mass in order to obtain a signal in the LIGO/VIRGO frequency band and stuck to the chirp masses 
of two typical neutron stars and a black hole pair respectively. 
The value of the graviton mass used in that cases exceeds the 
experimental upper limits on $m_g$ by many orders of magnitude. 
For the long dashed line we fix $m_g$ to its maximum value allowed by Solar System tests of 
gravity and adapt the chirp mass, which now has to be on the order 
of the mass of a galaxy. Thus we see already that the observed GW events are not easily 
explained in the context of the CGM with small graviton mass. But note that for a proper comparison 
with data we also have to consider the modified propagation of the gravitational wave signal in 
the CGM and thus the results shown in Fig.~\ref{fig:frequencies} would actually only be 
observed by a fiducial observer very close to the source, but not by us. 
Thus we work out the details of the wave form below.

For the massless mode of the gravitational wave $\tau = t_\mathrm{coal} - t = 
t_{\mathrm{coal},0} - t_0$, as for the large mass limit of conformal gravity.  
However,  in the case of small $m_g$ we find for the massive mode of the gravitational radiation 
that the modified speed of propagation must be taken into account when we 
evaluate what a distant observer sees.

\begin{figure}
\includegraphics[width = \linewidth]{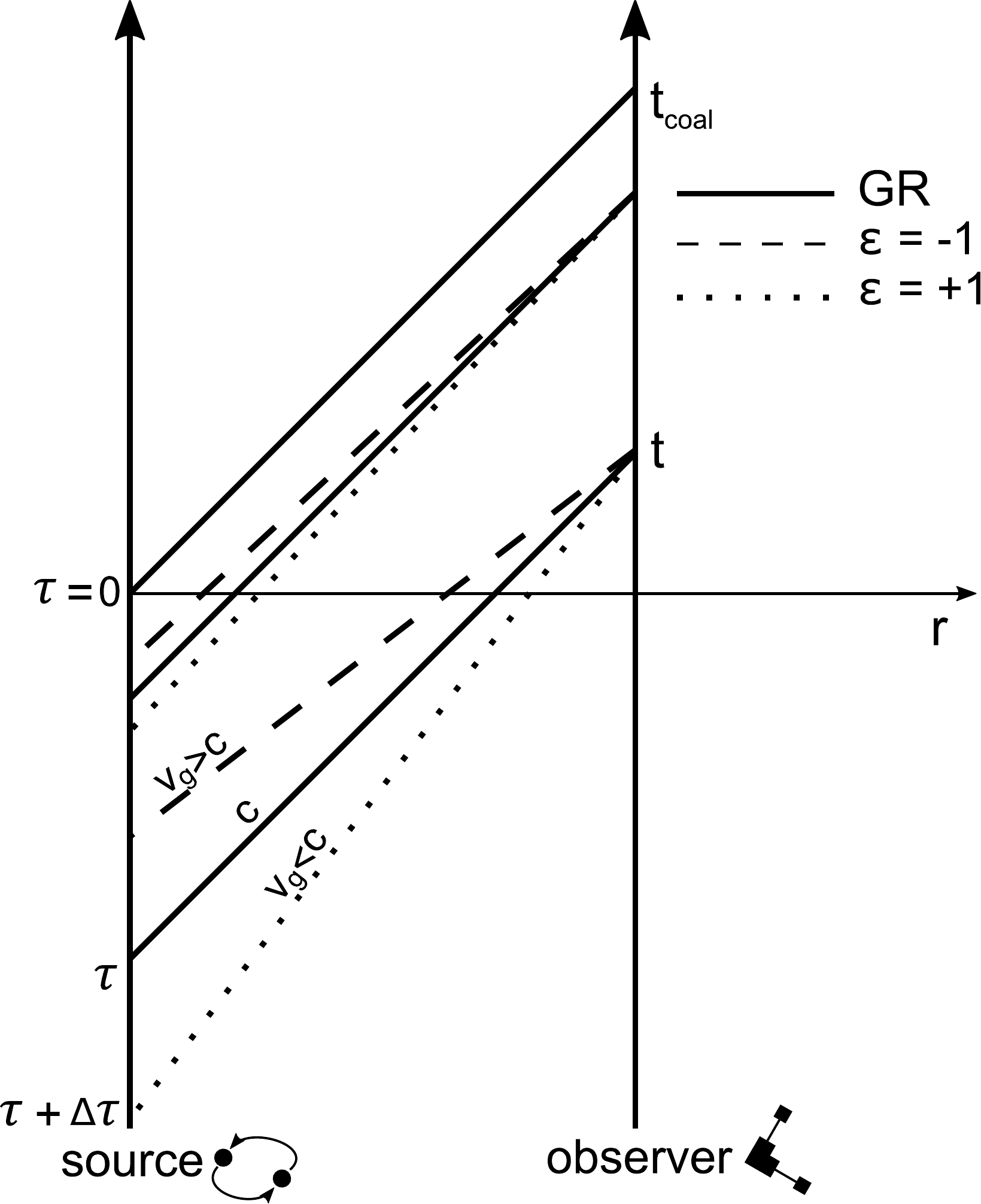}
\caption{World lines of massless and 
massive gravitational wave modes emitted by a binary source and observed by a detector.
The observer sees a superposition of massless modes emitted at $\tau$ and massive modes 
emitted at $\tau + \Delta \tau$. Note that $\Delta \tau > 0$ for $\epsilon = +1$ and 
$\Delta \tau < 0$ for $\epsilon = -1$. It is also shown that $\Delta \tau \to 0$ for $\tau \to 0$ meaning 
that the speed of the massive modes approaches the speed of light at coalescence.  
	\label{fig:traveltime}}
\end{figure}

Figure \ref{fig:traveltime} illustrates the situation. Well before coalescence at time $t$, the observer sees 
a superposition of the massless mode emitted at retarded time $t_0 = t - r/c$ and the massive mode
emitted at time $t_m = t - v_{g,\epsilon} r/c^2$. 
The time lag between the emission of the two modes is 
$\Delta\tau \equiv (1 - v_{g,\epsilon}/c)(r/c) \approx 
\epsilon (r/2c) (m_g c^2/\hbar \omega_\mathrm{gw})^2$, where we expanded the expression for the
group velocity, as the condition  $m_g c^2/\hbar \omega_\mathrm{gw} \ll 1$ holds in the small mass 
regime.

In the following we will assume that $\Delta \tau/\tau \ll 1$. This is an excellent assumption for 
the interesting parameter space as
\begin{align}
& \frac{\Delta \tau}{\tau} = \epsilon \frac{512}{5} \frac{r}{c} \left(\frac{m_g c^2}{\hbar}\right)^{\!\! 8} \!\!
 \left(\frac{G M_c}{c^3}\right)^{\!\! 5}\!\! \left(\frac{\tau}{5}\right)^{\! 2} \\
 & = \epsilon\, 4.24\! \times\! 10^{-66}\! \left[\frac{r}{\SI{1}{Gpc}}\right]\! \!
\left[\frac{m_g}{\SI{d-58}{kg}}\right]^{\! 8} \!\!
\left[\frac{M_c}{M_\odot}\right]^{\!5}\! \left[\frac{\tau}{\SI{1}{s}}\right]^{2}\!\!\!.
\end{align}
Since the value assumed for $m_g$ is an upper limit, and the value assumed for the distance of the 
source is of the order of the size of the observable universe, the condition is met for all chirp masses 
below $\SI{1.4d13}{} M_\odot$. It is even met for larger chirp masses if the source is located at distance 
$< 1$ Gpc. Note that the condition is not met for $m_g > \SI{2d-50}{kg}$, but those values are 
excluded based on Solar System tests.

The next step is to calculate the frequency at the time of the emission of the massive mode,  
\begin{equation}
 \omega_\mathrm{gw} (\tau + \Delta \tau) \approx  \omega_\mathrm{gw}(\tau) + \dot{\omega}_\mathrm{gw} \Delta \tau.
\end{equation}
Using (\ref{eqn:omegadotsmallmass}) we find 
\begin{equation}
 \omega_\mathrm{gw} (\tau + \Delta \tau) \approx  \omega_\mathrm{gw}(\tau) \left[ 1 - \epsilon 
 \frac{384}{125} \left[\frac{GM_c}{c^3}\right]^5  \left[\frac{m_g c^2}{\hbar}\right]^8 \frac r c \tau^2 \right].
\end{equation}

For the massless mode we obtain the phase by integrating \eqref{frequencyGW} from $\tau_i$ to 
$\tau$, 
\begin{align}
&\Phi(\tau) - \Phi_i = 
\int_{\tau_i}^{\tau} d\tau^\prime \omega_{\mathrm{gw}}(\tau^\prime) \nonumber\\
&= \frac{5}{16}\! \left(\frac{m_{g} c^2}{\hbar}\right)^{\!\! -3}\!\!\!  \left(\frac{GM_c}{c^3} \right)^{\!\! -5/2} \!\left[ \left(\frac{5}{\tau_i}\right)^{\!1/2}\!\!\!\! - \! \left(\frac{5}{\tau}\right)^{\!1/2} \right],
	\label{massless phase}
\end{align}
where $\Phi_i = \Phi(\tau_i)$.

For the massive mode we have to integrate \eqref{frequencyGW} from $\tau_i + \Delta \tau_i$ to 
$\tau + \Delta \tau$. Keeping only contributions that are linear in $\Delta \tau$, we find
\begin{align}
&\Phi(\tau + \Delta \tau) - \Phi(\tau_i + \Delta \tau_i) 
= \int_{(\tau + \Delta \tau)_i}^{\tau + \Delta \tau} d\tau^\prime \omega_{\mathrm{gw}}(\tau^\prime)\nonumber\\
&\approx \int_{\tau_i}^{\tau} d\tau^\prime \omega_{\mathrm{gw}}(\tau^\prime) + \omega_{\mathrm{gw}}(\tau) \Delta\tau - \omega_{\mathrm{gw}}(\tau_i) \Delta\tau_i \nonumber \\
& \approx \Phi(\tau) - \Phi_i \nonumber \\
& + \epsilon 16 \frac{r}{c} \left(\frac{m_g c^2}{\hbar}\right)^{\!\! 5}\!\!  
\left(\frac{GM_c}{c^3} \right)^{\!\!5/2} \!\!
\left[ \left(\frac{\tau}{5}\right)^{3/2} \!\! - \left(\frac{\tau_i}{5}\right)^{3/2}  \right].
\label{massive phase}
\end{align}

To find the waveform we start from Eqs.~\eqref{solution case1/3 comp 11}  
and \eqref{solution case1/3 comp 12} in which we have to replace 
$\omega_{\mathrm{gw}}t_0$ with $\Phi(\tau)$ and 
$\omega_{\mathrm{gw}}t_{\mathrm{m}}$ with $\Phi(\tau + \Delta \tau)$. After making use of 
Kepler's 3rd law \eqref{Kepler CG}, by which we eliminate the orbital radius, 
we must take care of the fact that the gravitational wave frequency 
has to be evaluated at different retarded times for the massless and massive modes, which 
yields
\begin{subequations}
\begin{align}
&h_{11}(t,r)  = -h_{22}(t,r) \nonumber\\
&= \! \frac{\epsilon}{4} \frac{c}{r}\! \left(\frac{\hbar}{m_{g}c^2}\right)^{\!\!2}\!
\left[ \frac{5}{\tau} \cos[\Phi(\tau)] - \frac{5}{\tau+ \Delta \tau} \cos[\Phi(\tau+ \Delta \tau)] \right]\!,
	\\
&h_{12}(t,r) = h_{21}(t,r)\nonumber\\
&=   \! \frac{\epsilon}{4} \frac{c}{r}\! \left(\frac{\hbar}{m_{g}c^2}\right)^{\!\!2}\!
\left[ \frac{5}{\tau} \sin[\Phi(\tau)] - \frac{5}{\tau+ \Delta \tau} \sin[\Phi(\tau+ \Delta \tau)] \right]\!.
\end{align}
\end{subequations}

Now we assume that the delay between the massive and massless modes 
is small and keep all terms linear in $\Delta \tau$. Let us demonstrate this for 
the $h_{11}$ component,
\begin{align}
& \left[\frac{1}{\tau} \cos[\Phi(\tau)] - \frac{1}{\tau+ \Delta \tau} \cos[\Phi(\tau+ \Delta \tau)] \right] \nonumber \\
& \approx \frac{\Delta\tau}{\tau} \left[ \sin[\Phi(\tau)] \Phi(\tau)^\prime  -  \frac{1}{\tau}\cos[\Phi(\tau)] \right].
\end{align}
With $\Phi(\tau)^\prime = \omega_{\mathrm{gw}}(\tau)$, we find
\begin{subequations}
\begin{align}
&h_{11}(t,r)  = -h_{22}(t,r) \nonumber\\
&\approx 4 \left(\frac{m_g c^2}{\hbar}\right)^{\!\! 3} \!\!
 \left(\frac{G M_c}{c^3}\right)^{\!\! 5/2}\!\! \left(\frac{\tau}{5}\right)^{\! 1/2} \times \nonumber \\
& \left[\sin[\Phi(\tau)]  -   \frac{32}{5}\! \left(\frac{m_{g} c^2}{\hbar}\right)^{\!\!3}\! 
\left(\frac{G M_c}{c^3}\right)^{\!\! 5/2}\!\! \left(\frac{\tau}{5}\right)^{\!\! 1/2}\!\! \cos[\Phi(\tau)] \right] ,
\label{sol:massive11} \\
&h_{12}(t,r) = h_{21}(t,r)\nonumber\\
&\approx  4 \left(\frac{m_g c^2}{\hbar}\right)^{\!\! 3} \!\!
 \left(\frac{G M_c}{c^3}\right)^{\!\! 5/2}\!\! \left(\frac{\tau}{5}\right)^{\! 1/2} \times \nonumber \\
& \left[- \cos[\Phi(\tau)]  -  \frac{32}{5}\! \left(\frac{m_{g} c^2}{\hbar}\right)^{\!\!3}\! 
\left(\frac{G M_c}{c^3}\right)^{\!\! 5/2}\!\! \left(\frac{\tau}{5}\right)^{\!\! 1/2}\!\! \sin[\Phi(\tau)] \right].
	\label{sol:massive12}
\end{align}
\end{subequations}

At leading order, the massless and massive mode cancel each other and 
the emission of gravitational waves is strongly suppressed when compared to GR or the large graviton mass situation. Another important observation is that the result does not depend on 
the distance of the observer from the source, which is a 
reflection of the Weyl symmetry of the model. For small graviton mass, the emitted 
gravitational waves are in the high energy regime of the theory where typically symmetries 
are seen more explicitly than at low energy scales. Moreover the result does not depend on the 
sign of $\epsilon$. And most importantly, the amplitude of the signal decreases towards 
coalescence and vanishes at $\tau = 0$, in stark contrast to the GR and large mass CGM
predictions.

To estimate the order of magnitude of the gravitational wave amplitude we can write the factor in 
front of the square bracket as 
\begin{align}
& 4 \left(\frac{m_g c^2}{\hbar}\right)^{\!\! 3} \!\!
 \left(\frac{G M_c}{c^3}\right)^{\!\! 5/2}\!\! \left(\frac{\tau}{5}\right)^{\! 1/2} = \nonumber \\
&  6.00 \times 10^{-35} \left(\frac{m_g}{\SI{d-58}{kg}}\right)^{\!\! 3} \!\!
 \left(\frac{M_c}{M_\odot}\right)^{\!\! 5/2}\!\! \left(\frac{\tau}{\SI{1}{s}}\right)^{\! 1/2}.
\end{align}
For the typically assumed chirp masses this is many orders of magnitudes smaller 
than the amplitude observed by the LIGO/VIRGO detectors. 
For chirp masses below the mass of a typical galaxy, the second terms in the 
square brackets are negligible and it is sufficient to compare LIGO/VIRGO observations to 
the leading terms for both polarizations.

Thus the predicted wave form for gravitational radiation from a coalescing binary for the CGM 
in the small graviton mass regime reads,
\begin{subequations}
\begin{align}
&h_{11}(t,r)  = -h_{22}(t,r) \nonumber\\
&\approx 4 \left(\frac{m_g c^2}{\hbar}\right)^{\!\! 3} \!\!
 \left(\frac{G M_c}{c^3}\right)^{\!\! 5/2}\!\! \left(\frac{\tau}{5}\right)^{\! 1/2} \sin[\Phi(\tau)], 
\label{sol:massive11final} \\
&h_{12}(t,r) = h_{21}(t,r)\nonumber\\
&\approx  - 4 \left(\frac{m_g c^2}{\hbar}\right)^{\!\! 3} \!\!
 \left(\frac{G M_c}{c^3}\right)^{\!\! 5/2}\!\! \left(\frac{\tau}{5}\right)^{\! 1/2} \cos[\Phi(\tau)].
	\label{sol:massive12final}
\end{align}
\end{subequations}
In the limit of vanishing $m_g$, there is no gravitational radiation emitted to the far 
field of the source.

\begin{figure}
\includegraphics[width = \linewidth]{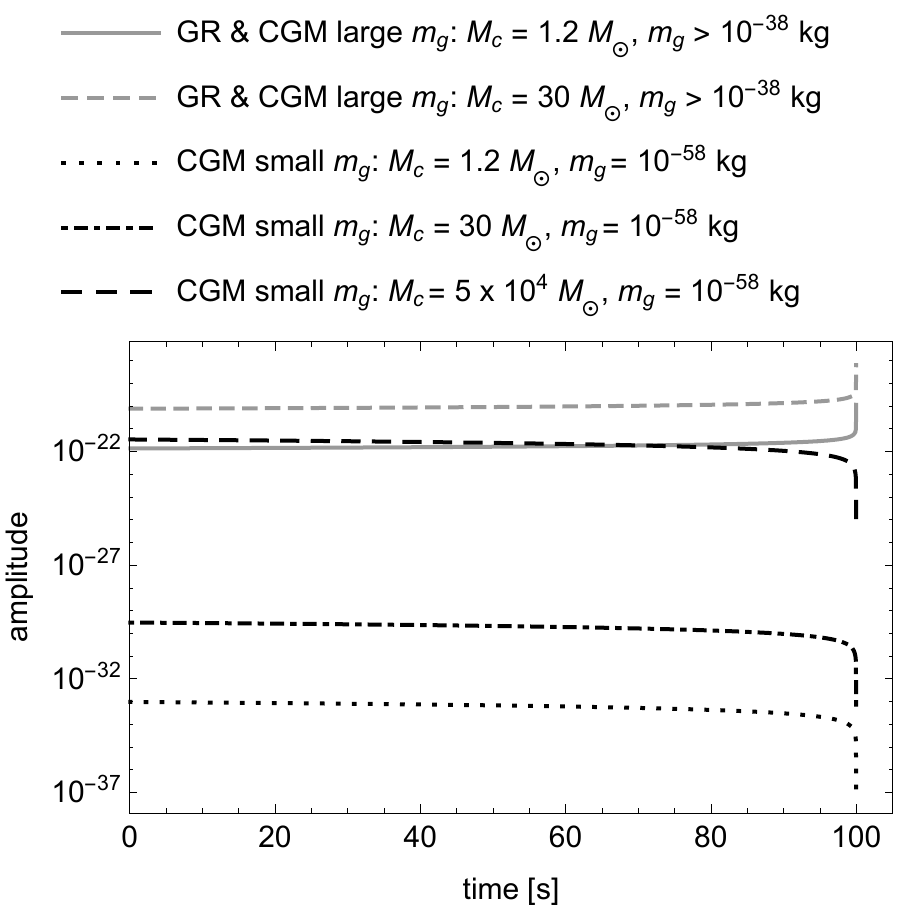}
\caption{Time evolution of the gravitational wave amplitudes during the last 100 
seconds before coalescence. The grey solid and dashed lines show the 
GR prediction for two different chirp masses at an assumed distance of $10$ Mpc. 
The GR prediction is identical to the CGM prediction with a large graviton mass. 
The black lines show three examples for the CGM with a small graviton mass with the maximally allowed 
value of the graviton mass. \label{fig:amplitudes}}
\end{figure}

In Fig. \ref{fig:amplitudes} we show the time evolution of the GW amplitudes for GR or 
the CGM with 
large $m_g$ (grey lines) and for the CGM for $m_g = 10^{-58}$ kg (black lines). The two GR examples 
are located at a distance of 10 Mpc and show chirp masses of $1.2 M_\odot$ and $30 M_\odot$.
The CGM results are valid for arbitrary distance. 
For comparison we show the chirp masses typical for the observed GW events in the context 
of GR as dotted and dashed-dotted lines. They give rise to amplitudes that are many orders of magnitude below those discovered by the LIGO/VIRGO  collaboration.  In order to match the 
typical amplitudes found, one would require a chirp mass of $5 \times 10^4 M_\odot$, which in turn 
would predict a gravitational wave frequency $\approx  3 \times 10^{21}$ Hz $(1$ s$/\tau)^{3/2}$, 
which is not at all in the frequency band. It is not possible to find values of $m_g$ and $M_c$ 
that fit both the frequency and amplitude range of the LIGO/VIRGO detectors and evolve on 
a typical timescale of a fraction of a second to about a few minutes.

\section{Conclusion \label{conclusion}}

We have calculated the effect of gravitational radiation on the orbit of a binary 
system of compact objects in the late inspiral phase in a conformal gravity model for 
large and small graviton masses. For the binary system we used the center of mass frame on a 
quasicircular orbit and in the Newtonian limit, which is justified because in the 
early phase of observed merger events the two bodies are still quite far 
apart. Furthermore, binary systems, although they may posses large eccentricity in the early 
phase of the inspiral, tend to circularize in the late inspiral phase very rapidly.

This work builds on the results obtained in \citep{2018_Caprini_AstrophysicalGravitationalWavesinConformalGravity_apa}, 
where the decrease of the orbital period of stellar binary systems in the CGM in the 
early stationary inspiral phase was investigated. 

The conformal models of gravity studied in this work allow for seven radiative degrees of 
freedom. 
Two of them are massless and behave very similar as in general relativity. The other five degrees of
freedom are massive. For a conserved energy-momentum tensor of the source only two of those 
five degrees of freedom can be excited by a binary system 
\citep{2018_Caprini_AstrophysicalGravitationalWavesinConformalGravity_apa}. These two modes 
propagate to a distant detector if the frequency of the gravitational wave is above 
the characteristic frequency that corresponds to the mass $m_g$. We referred to this as the 
small graviton mass case.

For large graviton mass we found that the observed wave form 
is (at least at leading order in a post-Newtonian expansion) indistinguishable from the 
GR result. Corrections from higher-derivative contributions 
of the CGM become important only on microscopic 
scales, which are irrelevant on the distance scales of binary systems. All modifications to the GR 
results are negligible, since they are much smaller than the error of measurement.

In the case of a small graviton mass the GW solutions look very different than in GR. In the 
parameter regime with $\Delta\tau/ \tau \ll 1$ the amplitude is independent of the 
distance to the binary system and is decreasing as coalescence is approached. 
We conclude that for small graviton mass both regimes of the studied CGM 
($\epsilon = \pm 1$) cannot explain the observed gravitational wave events.

Suppressed GWs have also been found in a version of torsionfree gravitational 
Yang-Mills gauge theories based on the SO(4,2) conformal group of the Minkowski spacetime \cite{2016GegenbergInfraredmodificationgravity,2017GegenbergBigbounceslow}, which coincides with the considered CGM for $\epsilon = +1$. The retarded Green's function in a de Sitter vacuum derived in \cite{2018GegenbergGravitationalwavedefocussing} resembles our results in flat spacetime, which we found in \cite{2018_Caprini_AstrophysicalGravitationalWavesinConformalGravity_apa}.

Our new findings, together with the results from 
\cite{2018_Caprini_AstrophysicalGravitationalWavesinConformalGravity_apa}, restrict 
the CGM to large graviton masses and the regime $\epsilon = +1$, 
which cannot explain galaxy rotation curves without dark matter. 
The large graviton masses give rise to a modification of GR at high energies and small distances 
and therefore the CGM remains an interesting target for further studies.

\begin{acknowledgments}
We wish to thank Chiara Caprini and Felipe F. Faria for valuable comments and discussions. We are grateful to Mariafelicia De Laurentis, Sanjeev Seahra, Jack Gegenberg and Sumanta Chakraborty for useful comments. PH and DJS acknowledge financial support from Deutsche
Forschungsgemeinschaft (DFG) under Grant No. RTG 1620 "Models of Gravity".
We also thank the COST Action CA15117 "Cosmology
and Astrophysics Network for Theoretical Advances and Training Actions (CANTATA)", 
supported by COST (European Cooperation in Science and Technology).
\end{acknowledgments}

\appendix

\section{\label{appendix A}Conventions}

The signature of the metric is 
$g = \text{diag}\left(-,+,+,+\right)$.
The Riemann tensor in terms of Christoffel symbols is given by
\begin{equation}
R_{\mu\nu\kappa}^{\lambda} = 
- \left(\partial_{\nu}\Gamma_{\mu\kappa}^{\lambda} - 
	\partial_{\kappa}\Gamma_{\mu\nu}^{\lambda} + 
	\Gamma_{\nu\alpha}^{\lambda}\Gamma_{\mu\kappa}^{\alpha} - 
	\Gamma_{\kappa\alpha}^{\lambda}\Gamma_{\mu\nu}^{\alpha}\right).
\end{equation}
From this we find the Ricci tensor $R_{\mu\kappa} = g^{\lambda\nu} R_{\lambda\mu\nu\kappa}$ and the Ricci scalar $R = g^{\mu\kappa} R_{\mu\kappa}$. 
The Weyl tensor is given by the expression 
\begin{widetext}
\begin{equation}
C_{\lambda\mu\nu\kappa} = R_{\lambda\mu\nu\kappa}+\frac{1}{6} R \left[g_{\lambda\nu}g_{\mu\kappa}-g_{\lambda\kappa}g_{\mu\nu}\right] -\frac{1}{2}\left[g_{\lambda\nu}R_{\mu\kappa}-g_{\lambda\kappa}R_{\mu\nu}-g_{\mu\nu}R_{\lambda\kappa}+g_{\mu\kappa}R_{\lambda\nu}\right].
	\label{eq: Weyl Tensor} 
\end{equation}
The Einstein equations in the convention used in \citep{weinberg1972gravitation} reads 
\begin{equation}
G_{\mu\nu} \equiv R_{\mu\nu} - \frac{1}{2}g_{\mu\nu}R =
- 8\pi G T_{\mu\nu} + \Lambda g_{\mu\nu}.
\end{equation}
Finally, the Bach tensor can be written as 
\begin{equation}
W^{\mu\nu} = -\frac{1}{6} g^{\mu\nu} R_{\;\;;\beta}^{;\beta}+R_{\;\;\;\;\;\;\;;\beta}^{\mu\nu;\beta} - R_{\;\;\;\;\;\;\;;\beta}^{\mu\beta;\nu} - R_{\;\;\;\;\;\;\;;\beta}^{\nu\beta;\mu} - 2R^{\mu\beta}R_{\:\,\beta}^{\nu} + \frac{1}{2}g^{\mu\nu}R_{\alpha\beta}R^{\alpha\beta} + \frac{2}{3}R^{;\mu;\nu} + \frac{2}{3}RR^{\mu\nu} - \frac{1}{6}g^{\mu\nu}R^{2}.
	\label{eq: Bach Tensor}
\end{equation}
\end{widetext}

\begin{thebibliography}{48}
\bibitem [{sim()}]{simplifiednotation}%
  \BibitemOpen
  \href@noop {} {\bibinfo  {journal} {We have used simplified notation. The
  kinetic part of the fermionic energy-momentum tensor is given by
  $i\bar{\psi}\gamma^{\mu}(x)\left[\partial_{\mu} + \Gamma_{\mu}(x)\right]\psi
  - i\bar{\psi}\left[\overleftarrow{\partial_{\mu}} + \Gamma_{\mu}(x)\right]
  \gamma^{\mu}(x)\psi$ in order to be hermitian}\ }\BibitemShut {NoStop}%
\end{thebibliography}
\end{document}